\documentclass[conference]{IEEEtran}
\IEEEoverridecommandlockouts
\usepackage{cite}
\usepackage{svg}
\usepackage{amsmath,amssymb,amsfonts}
\usepackage{graphicx}
\usepackage{textcomp}
\usepackage{xcolor}
\usepackage{tabularx}
\usepackage{makecell}
\def\BibTeX{{\rm B\kern-.05em{\sc i\kern-.025em b}\kern-.08em
  T\kern-.1667em\lower.7ex\hbox{E}\kern-.125emX}}
\usepackage{booktabs}
\usepackage{algpseudocode}
\usepackage{amsmath}
\usepackage{amssymb}
\usepackage{multirow}
\usepackage{subcaption}
\usepackage{hyperref}
\begin{document}

\title{Exploring Algorithmic Explainability: Generating Explainable AI Insights for Personalized Clinical Decision Support Focused on Cannabis Intoxication in Young Adults}
% Exploring Algorithmic Explainability of Cannabis-Intoxicated Behaviors Towards Human-AI Decision-Making\\

% Understanding Algorithmic Explainability in cannabis-Intoxicated Behaviors: Implications for Human-AI Decision-Making

% \title{Unlocking Personalized Interventions: A Case Study on cannabis-Intoxicated Behaviors in Young Adults with Explainable AI Insights for Clinical Decision Support\\

\author{\IEEEauthorblockN{Tongze Zhang \href{https://orcid.org/0000-0002-3375-7136}{\includegraphics[scale=0.06]{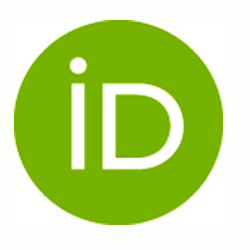}}}
\IEEEauthorblockA{%\textit{Stevens Institute of Technology} \\
\textit{Stevens Institute of Technology}\\
Hoboken, New Jersey }
\and

\IEEEauthorblockN{Tammy Chung \href{https://orcid.org/0000-0002-1527-2792}{\includegraphics[scale=0.06]{figures/orcid.png}}}
\IEEEauthorblockA{%\textit{dept. name of organization (of Aff.)} \\
\textit{Rutgers University}\\
Newark, New Jersey }
\and
\IEEEauthorblockN{Anind Dey \href{https://orcid.org/0000-0002-3004-0770}{\includegraphics[scale=0.06]{figures/orcid.png}}}
\IEEEauthorblockA{%\textit{dept. name of organization (of Aff.)} \\
\textit{University of Washington}\\
Seattle, Washington}
\and
\IEEEauthorblockN{*Sang Won Bae \href{https://orcid.org/0000-0002-2047-1358}{\includegraphics[scale=0.06]{figures/orcid.png}}}
\IEEEauthorblockA{%\textit{Stevens Institute of Technology} \\
\textit{Stevens Institute of Technology}\\
Hoboken, New Jersey}
% \and
}

\maketitle

\begin{abstract}
\textcolor{black}{
As an increasing number of states adopt more permissive cannabis regulations, the necessity of gaining a comprehensive understanding of cannabis’s effects on young adults has grown exponentially, driven by its escalating prevalence of use. By leveraging popular eXplainable Artificial Intelligence (XAI) techniques such as SHAP (SHapley Additive exPlanations), rule-based explanations, intrinsically interpretable models, and counterfactual explanations, we undertake an exploratory but in-depth examination of the impact of cannabis use on individual behavioral patterns and physiological states. \textcolor{black}{This study explores the possibility of facilitating algorithmic decision-making by combining interpretable artificial intelligence (XAI) techniques with sensor data, with the aim of providing researchers and clinicians with personalized analyses of cannabis intoxication behavior. SHAP analyzes the importance and quantifies the impact of specific factors such as environmental noise or heart rate, enabling clinicians to pinpoint influential behaviors and environmental conditions. SkopeRules simplify the understanding of cannabis use for a specific activity or environmental use. Decision trees provide a clear visualization of how factors interact to influence cannabis consumption. Counterfactual models help identify key changes in behaviors or conditions that may alter cannabis use outcomes, to guide effective individualized intervention strategies.} This multidimensional analytical approach not only unveils changes in behavioral and physiological states after cannabis use, such as frequent fluctuations in activity states, nontraditional sleep patterns, and specific use habits at different times and places, but also highlights the significance of individual differences in responses to cannabis use. These insights carry profound implications for clinicians seeking to gain a deeper understanding of the diverse needs of their patients and for tailoring precisely targeted intervention strategies. Furthermore, our findings highlight the pivotal role that XAI technologies could play in enhancing the transparency and interpretability of Clinical Decision Support Systems (CDSS), with a particular focus on substance misuse treatment. This research significantly contributes to ongoing initiatives aimed at advancing clinical practices that aim to prevent and reduce cannabis-related harms to health, positioning XAI as a supportive tool for clinicians and researchers alike.}

% In this case study, we leverage Explainable Artificial Intelligence (XAI) models, specifically SHAP and LIME, augmented by counterfactual explanations, to conduct an in-depth exploration of the multifaceted consequences of cannabis use behaviors in young adults. Our study employs XAI techniques to conduct a meticulous analysis of the impact of cannabis use on both individual behavioral patterns and physiological states. Through our rigorous analysis, we elucidate substantial shifts in user behavior and physiological responses, encompassing frequent fluctuations in activity status, unconventional sleep patterns, and usage habits specific to particular times and locations. 
\end{abstract}


\section{Introduction}

The legalization of cannabis for recreational use in 18 states, \textcolor{black}{as well as Washington, DC}, and for medicinal use in 37 states has led to increased accessibility, especially among young adults \cite{olfson2018cannabis}. This raises concerns about potential negative health impacts of cannabis. Given the potential risks associated with use and the increased availability of cannabis\textcolor{black}{\cite{NIH2021}}, helping individuals to become more aware of the effects of cannabis in real-time could prevent cannabis-related negative consequences.

\textcolor{black}{Thanks to state-of-the-art AI (artificial intelligence) technologies}, mobile and wearable sensors have been developed, now enabling the possibility of detecting binge drinking and high-risk drinking \cite{bae2023leveraging, bae2017detecting, bae2018mobile}; and cannabis intoxication \cite{Bae2023, bae2021mobile}. This real-time and context-rich data collection from these sensors is groundbreaking, as detecting individuals’ behaviors in everyday settings provides opportunities to predict intoxicated behaviors associated with substance use. \textcolor{black}{In the face of rising substance misuse and the challenges posed by increased accessibility of substances like cannabis, advances in technology, notably AI and wearable sensors, offer innovative solutions that could support}, for example, Just-In-Time interventions to prevent substance use-related negative consequences. A critical barrier in using the results of AI technologies to inform intervention development is the lack of transparency regarding "how" the model makes its prediction. That is, the models themselves do not reveal the most informative features that contribute to model performance. Thus, the models predict outcomes, but are limited in guiding the development of intervention content based on their most informative features. However, the growing transparency in AI approaches addresses the "black box" problem, and can help identify potentially important correlates that inform prediction, and that can guide intervention development. By elucidating how the AI model generates decisions, clinicians can discover the logic in the AI model's decision making, obtain new knowledge derived from validated model features; gain increased algorithmic transparency \cite{pierce2022explainability}, mitigate heuristic bias \cite{ali2023explainable}, and obtain insight \cite{liao2020questioning} in designing interventions based on model decision making processes. The contributions of this study are multifaceted. We propose the value of using individual-level Explainable Artificial Intelligence (XAI), utilizing popular local explanation models such as SHAP (SHapley Additive exPlanations) \cite{lundberg2020local}\cite{lundberg2017unified}, counterfactual \cite{wachter2017counterfactual}, leveraging SkopeRules \cite{skope_rules} and Intrinsically Interpretable Models \cite{loh2011classification}, to understand how and why our trained algorithm predicts cannabis intoxication behaviors. 
%\textbf{By elucidating how the AI model generates decisions, clinicians can discover the logic in the decision-making of the AI model, could obtain new knowledge derived from validated values; increased algorithmic transparency , mitigate heuristic bias , and gain insight in designing interventions based on model decision-making processes. }

SHAP values increase transparency by quantifying the impact of each characteristic on predicted outcomes. By detailing the impact of different factors such as sleep patterns, behavioral changes, and duration of cannabis use on intoxication, the decision-making process of the model can be understood. SkopeRules extracts rules from categorized data. SkopeRules combines the advantages of decision trees and rule learning to generate easy to understand lists of decision rules. These rules can guide clinicians in identifying key factors that contribute to an individual’s cannabis intoxicated behaviors. The intrinsically interpretable tree model visualizes the decision making process, which could make it easier for clinicians and patients to understand the logic of the predictions. This clarity could also increase trust in the recommendations of the AI system. Counterfactual explanations provide personalized insights into the behavioral and physiological conditions that strongly impact cannabis intoxication by showing how changing specific characteristics would alter predictions. 
Extracting interpretable rules linking behavioral and physiological states to cannabis use could provide information associated with decision rules, alternatives and hypothetial explanations XAI models generate, which could help clinicians understand possible scenarios that might be adjusted. 

% facilitate the design of interventions. 

Tailoring models to individual patients and using XAI to interpret these models ensures that decisions are based on personalized data rather than generalized assumptions. Using a combination of different XAI techniques provides a multidimensional understanding of the effects of cannabis. This approach helps distill complex model predictions into comprehensible knowledge, enabling clinicians to develop targeted interventions. XAI can provide actionable information that enables clinicians to guide their decisions and develop personalized strategies based on physiological and behavioral data. Through interpretive analysis using XAI frameworks, this study demonstrates the impact of cannabis use on individual behavior, physiological states, and daily habits, providing clinicians with information to understand and improve patient care.
\textcolor{black}{
Therefore, this case study poses a critical question: Can algorithmic decisions, facilitated by various AI explanations integrated with passive sensing data, potentially help researchers and clinicians provide actionable information about cannabis-intoxicated behaviors? If successful, this study would provide an advance in machine learning in personalized healthcare through the innovative application of XAI technology to understand the impact of cannabis use on behavior. Clinical decision-making would be enhanced by integrating XAI into a clinical decision support system, resulting in increased transparency and trust in machine learning models and potential to support tailored healthcare for patients .
}
%If successful, insights from this study may lead to informed decisions and strategies for designing personalized interventions. 
 %We aim to navigate the potential and possibilities of integrating XAI into a decision-making process that could support, for example, Just-In-Time intervention. Through this integration, we will provide insights to better understand individual-level substance use behavior that could influence the design of personalized intervention systems. 

\section{Background Literature and Related Work}
\subsection{Mobile Sensing in Substance Use Behavior }
Mobile sensing and predictive modeling have emerged as powerful methods \textcolor{black}{to detect and manage substance misuse} \cite{liu2021digital}\cite{tofighi2018role}\cite{hayes2022using}. These methods offer early detection capabilities through real-time monitoring and the analysis of behavioral patterns to aid in identifying signs of substance misuse. Furthermore, they enable personalized treatment by providing interventions matched to an individual's specific behavioral patterns and triggers \cite{carreiro2015imstrong}. A 2018 review suggests that technology-based tools can effectively address substance misuse, providing evidence-based interventions and typically rely on a single device or data source for monitoring and analysis. This reliance on a single device may lead to an incomplete understanding of behavioral patterns, limiting the effectiveness of early detection and personalized interventions. To overcome this limitation, research is needed to explore the use of multiple devices and data sources. By integrating data from multiple sources, such as smartphones, wearables, and other devices, a more comprehensive view of behavioral patterns can be obtained, thereby potentially improving the accuracy and efficiency of substance use detection and management.
\textcolor{black}{
The integration of AI interpretation with sensing data remains underexplored in the field of personalized healthcare and substance use research. This study contributes to filling this gap by exploring the potential roles of XAI techniques to understand if they could be used to obtain meaningful and actionable insights into an individual’s cannabis-intoxicated behavior. The application of XAI to clinical decision support systems represents an advancement in the field, providing a patient-centered approach to healthcare.
}

\subsection{Clinical Decision Support Systems (CDSS) Exist for Substance Misuse, but are Understudied for Early Intervention Substance Use Behavior}
Clinical decision support systems (CDSS) are computerized tools that help healthcare providers make better decisions by providing clinical knowledge, patient information, and other health data \cite{taheri2021effects}\cite{bian2022using}\cite{nishimura2022toward}. In the field of substance use disorders, CDSS have been developed to address the opioid epidemic, fentanyl-related overdoses, and addiction. For instance, the National Institute on Drug Abuse created a decision support tool integrating electronic technology into EHRs (Electronic Health Records) to enhance opioid use disorder screening and treatment in primary care settings \cite{bart2020developing}. CDSS enhance patient safety, clinical management, and cost-efficiency, but face challenges like alert fatigue, workflow disruptions, and data quality concerns \cite{sutton2020overview}\cite{jia2016effects}\cite{kilsdonk2016uncovering}\cite{dowding2009nurses}\cite{ash2007some}\cite{khalifa2016improving}. These systems can also disrupt the natural clinical workflow and increase cognitive load \cite{kilsdonk2016uncovering}\cite{dowding2009nurses}. Other challenges and limitations of CDSS include data quality, system maintenance, interoperability, and financial barriers \cite{sutton2020overview}. These issues can affect the accuracy, reliability, and validity of CDSS, as well as its adoption, implementation, and sustainability \cite{sutton2020overview}. Another important issue, as demonstrated by Demaraj et al., is that limited technological proficiency can create barriers when interacting with CDSSs \cite{devaraj2014barriers}\cite{leslie2006clinical}. This challenge can vary depending on the CDSS's design, with some being overly complicated and reliant on user expertise \cite{murray2011difficult}. To address this, CDSS should aim to align with the core functionalities of the existing system \cite{lai2006potential}. However, as all new systems involve a learning curve, it may be beneficial to assess users' technological competence initially and provide additional training to maximize CDSS utilization \cite{ojeleye2016ensuring}. 
Research on CDSS with young adults who report substance-related intoxication in daily life remains largely unexplored, especially when focusing on an individual level analysis of substance use.

% In this case study, our participants were young adults from the general community, not specifically individuals seeking treatment for substance use disorders. Their insights and experiences are likely more pertinent to digital interventions aimed at prevention or early intervention, rather than to clinical decision-support systems intended for individuals in addictions treatment.

\subsection{Explainable
Artificial Intelligence (XAI)}

In recent years, several methods have been proposed and widely used in different scenarios, including SHAP (SHapley Additive exPlanations) \cite{lundberg2020local}\cite{lundberg2017unified}, LIME (Local Interpretable Model-Agnostic Explanations) \cite{ribeiro2016should}, and counterfactual explanations\cite{wachter2017counterfactual}. In the field of XAI, researchers have used a variety of methods to improve the understandability and transparency of models. These methods can be broadly categorized into \textcolor{black}{three} groups: Feature interaction and importance, knowledge distillation and rule extraction, intrinsically interpretable models \cite{payrovnaziri2020explainable}. Please refer to Table \ref{table:xai_tools}.

\textcolor{black}{
Feature interaction and importance: this category covers how to interpret and understand the dependence of model predictions on input features. It includes techniques such as SHAP%(SHapley Additive exPlanations)
, which assigns the degree of contribution of each feature to model predictions by means of game-theoretic Shapley values; LIME%(Local Interpretable Model-agnostic Explanations)
, which explains individual predictions by means of locally approximated models to explain individual predictions; Accumulated Local Effects (ALE) \cite{apley2020visualizing}, which is used to assess the local variation of features to predictions; and Anchor \cite{ribeiro2018anchors}, which provides model predictions by determining the combination of features necessary to keep predictions stable and provide an "anchoring".}

\textcolor{black}{
Knowledge distillation and rule extraction: this category makes the decision-making process of a complex model transparent by transforming its decision logic into a more understandable form, such as a rule or a simplified model. SkopeRules \cite{skope_rules} provide a method for identifying and extracting the most important decision rules of a predictive model, while Counterfactual \cite{wachter2017counterfactual}\cite{guidotti2022counterfactual}\cite{shang2022not} interpretation helps to understand the decision boundaries of the model by modifying the input data to change the predictions.}

\textcolor{black}{
Intrinsically interpretable models: This category emphasizes the development of intrinsically highly interpretable models \cite{loh2011classification}, such as decision trees \cite{scikit-learn}, linear models \cite{scikit-learn}, whose decision-making processes are directly interpretable without the need for additional layers of interpretation.}

\textcolor{black}{
In recent years, the field of XAI has made significant progress, especially in explaining the decision-making process of machine learning models. These advances come from theoretical and empirical studies from different disciplines, aiming to meet the growing demand for high-quality human-computer interactions. Payrovnaziri et al. (2020) \cite{payrovnaziri2020explainable} demonstrated the use of XAI models in biomedical applications through a systematic scoping review of real-world EHR data. From a medical professional's perspective, the XAI approach has both potential benefits and challenges. Lauritsen et al. (2020) \cite{lauritsen2020explainable} presented an interpretable AI early warning score system (XAI-EWS) for the prediction of acute critical conditions. The system emphasizes the importance of providing explanations in a clinical setting and how relevant EHR data can be used to aid in clinical decision-making. Liao et al. (2020) \cite{liao2020questioning} explore XAI from the perspective of user experience and design practices through interviews with user experience and design practitioners. These studies reveal that while XAI has great potential to provide medical decision support, its evaluation in the medical field has not yet been fully and formally practiced \cite{liao2020questioning}. There is a gap between existing research and the practice of creating interpretable AI products. This is especially true in terms of how XAI algorithms can be effectively integrated into user-friendly product designs and how results can be interpreted \cite{payrovnaziri2020explainable}. }

% \textcolor{black}{
% Data dimensionality reduction: data dimensionality reduction techniques such as PCA (Principal Component Analysis) \cite{holland2008principal} and t-SNE (t-distributed Stochastic Neighbor Embedding) \cite{zhou2018t} are often used to visualize high-dimensional datasets and help understand the intrinsic structure of the data, although they do not directly provide by themselves, the interpretation of model predictions.}

% \begin{figure}[htp]
%  \centering
%  \includegraphics[width=0.8\linewidth]{figures/XAI-1.png}
%  \caption{Overview of ML integrated Explainable AI pipeline }
%  % \Description{}
%  \label{img:pipline}
% \end{figure}

\begin{table*}[htbp]
\centering
\caption{Underlying principles of various XAI tools}
\label{table:xai_tools}
\begin{tabular}{|c|c|c|}
\hline
\textbf{XAI/ML Methods} & \textbf{Algorithmic Decision} & \textbf{Underlying Principle} \\ \hline
\thead{SHapley Additive exPlanations\\(SHAP)} & \thead{Feature Interaction and\\Importance categorization} & \thead{Cooperative game theory - assigning each feature \\an importance value for a particular prediction \cite{lundberg2017unified}.} \\ 
% \hline
% Counterfactual & \thead{Knowledge Distillation\\and Rule Extraction} & \thead{Understanding model predictions by \\exploring alternative scenarios where input features are modified \cite{wachter2017counterfactual}.} \\ 
\hline
% Tree model & Intrinsically Interpretable Models & \thead{Hierarchy of decisions - using a tree-like model of decisions \\and their possible consequences \cite{loh2011classification}.} \\ 
 % \hline
SkopeRules & \thead{Knowledge Distillation\\and Rule Extraction} & \thead{Rule-based induction - extracting \\if-then rules from data based on feature conditions \cite{dembczynski2008maximum}.} \\ 
\hline
Tree model & Intrinsically Interpretable Models & \thead{Hierarchy of decisions - using a tree-like model of decisions \\and their possible consequences \cite{loh2011classification}.} \\ 
\hline
Counterfactual & \thead{Knowledge Distillation\\and Rule Extraction} & \thead{Understanding model predictions by \\exploring alternative scenarios where input features are modified \cite{wachter2017counterfactual}.} \\ 
\hline
\end{tabular}
\end{table*}

\subsection{The Role of XAI in CDSS }

XAI enhances the interpretability of AI systems, providing insights into their general and specific predictions \cite{morrison2023evaluating}\cite{vellido2020importance}. By using interpretable AI to analyze behavioral changes due to cannabis use, more accurate and personalized data can be provided to support CDSS in making better recommendations when making clinical decisions. One of the main benefits of XAI is providing transparency into the model's decision-making process. When it comes to medical decision-making, understanding how the model arrived at a particular recommendation is critical to building trust between physicians and patients. This can help physicians better understand the recommendations provided by a CDSS and may increase the likelihood that they will follow them. Notably, LIME offers local explanations \cite{adadi2018peeking}, SHAP evaluates feature influence on predictions \cite{lundberg2020local}, and counterfactual explanations present hypothetical scenarios showing alternate model predictions \cite{wachter2017counterfactual}. In CDSS, XAI enhances reliability, safety, and trust by clarifying machine learning decisions, aiding in error detection, and fostering improved doctor-patient communication \cite{kunapuli2018decision}\cite{liao2020questioning}\cite{vorm2018assessing}\cite{xie2020chexplain}\cite{lamy2019explainable}\cite{arrieta2020explainable}\cite{schoonderwoerd2021human}\cite{luz2020machine}. XAI visualizations enhance AI model transparency, helping patients and clinicians understand and trust model decisions, leading to greater acceptance of results \cite{bussone2015role}\cite{adadi2018peeking}.

User acceptance and trust are essential for the effective deployment of CDSS in healthcare. However, the decision-making process of many advanced CDSS, especially those based on machine learning, is often not transparent to the user and operates as a "black box" \cite{nasir2023ethical}. This lack of transparency means that even if a system is able to make decision recommendations, healthcare professionals may be skeptical of those recommendations if the logic and rationale behind them are not clearly explained through various XAI methods \cite{amann2020explainability}. Furthermore, effective operation and understanding of CDSS may require a certain level of technical knowledge that not all healthcare professionals possess. To optimize this situation, it is critical to increase the transparency and interpretability of the CDSS decision-making process, as well as to lower the technical threshold for operating the system \cite{schwartz2022factors}. This will not only help to increase healthcare professionals' trust in the systems' recommendations, but can also promote their wider acceptance and use of these systems. By implementing a user-friendly interface design and providing easy-to-understand explanations of decisions \cite{kanstrup2011four}, CDSS can be made a trusted assistant for healthcare professionals, further enhancing the quality of healthcare services and patient care outcomes.

% \textcolor{black}{
% Several studies not only confirm the importance of model interpretability and transparency in healthcare applications, but also exemplify the potential of XAI to bridge the gap between AI computational power and clinical utility. For example, one of the case studies discussed highlights how interpretable AI models (e.g., models that leverage LIME, SHAP, and counterfactual interpretation) can facilitate a nuanced understanding of behavioral change, especially in complex situations like mental health and substance use \cite{ehsan2023charting}. This coincides with our advocacy for an integrated approach to CDSS in which the full potential of XAI to improve reliability, safety, and trust is realized. By capitalizing on XAI's documented strengths in improving error detection, patient-physician communication, and overall decision-making reliability \cite{lai2021towards}.
% }

\subsection{Personalized Intervention in Substance Misuse}
With an understanding of the behavioral changes that occur as a result of cannabis use, CDSS can use this information to provide patients with more individualized and targeted recommendations for interventions that will improve treatment outcomes. The field of personalized medicine has made significant progress in recent years, with many studies focusing on optimizing treatments by taking into account the unique characteristics of individuals. Personalized trials are widely recognized as an effective method of collecting large amounts of data through crossover designs and controlled interventions to determine whether a treatment is effective for an individual. This approach not only avoids the use of ineffective or harmful drugs, but also improves treatment outcomes and patient satisfaction \cite{olsson2005one}\cite{schork2019artificial}. In the field of neurodegenerative diseases, researchers have proposed personalized prevention and treatment strategies. Lifestyle modifications, such as staying cognitively active, avoiding loneliness and depression, eating a balanced diet, and supplementing with nutrients, are considered effective ways to reduce the risk of Alzheimer's disease. The use of precision medicine in the treatment of neurodegenerative diseases has also received attention \cite{peng2016towards}. By utilizing genomic, proteomic, and other multi-omics data, researchers are able to achieve early diagnosis, risk assessment, intervention selection, and affect evaluation, emphasizing the concepts of individualized, holistic, and preventive medicine. The application of this approach is not limited to drug therapy, but also includes lifestyle and behavioral interventions, as well as consideration of environmental and social factors \cite{goetz2018personalized}.

\section{Method}

\subsection{Data Preparation}

Data were collected via a mobile sensing app, the AWARE framework \cite{ferreira2015aware}, which was designed to continuously collect passive sensor data from smartphones, shedding light on young adults' behavioral patterns like location, physical movement, and device usage. This app, available on both Android and iOS, monitored 102 unique smartphone sensor, such as GPS and accelerometer data. In parallel, participants wore a Fitbit Charge 2, capturing physiological data like heart rate and step count, hypothesized to signal episodes of acute cannabis intoxication. In total, we collected 1,556 reports indicating no cannabis use and 221 reports of cannabis use from 57 \textcolor{black}{young adults (ages 18-25) recruited from the community who reported current cannabis use (primarily smoke or vape). We excluded participants with missing Fitbit data or who did not report cannabis not-intoxication during the study period. We focused on data from 34 participants who reported both cannabis use and non- use. From this subset, we obtained 772 reports, with 132 indicating self-reported cannabis-induced “high” or acute intoxication and 640 showing either no cannabis consumption or no resulting intoxication.}

% For most of the sensor features, we calculated their minimum, maximum, mean, median, and standard deviation. We chose a 5-minute window of time to extract sensor feature statistics because research has shown that heart rates increase significantly after 2-3 minutes of smoking cannabis . We obtained heart rate, sleep, and step data using Fitbit. 

\textcolor{black}{
For most of the sensor features, we calculated their minimum, maximum, mean, median, and standard deviation. We chose a 5 minute window of time to extract sensor feature statistics because lab research has shown that heart rates increases significantly roughly after 2-3 minutes after starting to smoke cannabis \cite{zuurman2009biomarkers}. We obtained heart rate, sleep, and step count data from the Fitbit.
}
For heart rate analysis, we further analyzed heart rate characteristics such as kurtosis (peaks) and skewness (asymmetry) \cite{groeneveld1984measuring}, which may provide insight into cannabis-induced physiological changes. We focused on events that lasted no longer than 3 hours and ignored data within 3 hours of another smoking episode because of concerns about how the effects of cannabis might continue between episodes. Additionally, to account for possible delays in self-reporting, we omitted the 30 minutes prior to the start time of each report. We divided the data into 5-minute segments labeled into two categories: 'not intoxication', label is 0, and ’intoxication’, label is 1.

% Analyses included 34 participants aged between 18 and 24, with a mean age of 19.6 and a standard deviation of 1.7 years. Females represented 61.8\% and males 38.2\%. The majority 70.6\% self-identified as White, 14.3\% Black, 11.4\% Asian, and 5.7\% as other race. Age of first cannabis use was, on average, at 16.5 years, with regular use starting, on average, at 17 years. Roughly one-third (35.3\%) reported daily cannabis usage. 

\subsection{Approach and Rationale for Participant Selection}
A representative sample selection methodology was used to analyze the data. This method aims to accurately differentiate between different levels of impact after cannabis use in order to provide a more nuanced understanding of how cannabis affects users. 

We categorized users into two groups based on self-reported levels of feeling high during cannabis use: Low intoxicated users, with high levels less than or equal to 3, and "moderate and intoxicated users", with high levels greater than 3. The purpose of employing this stratified sample selection method was to capture and understand the unique characteristics of the varying intensities of impacts during cannabis use. By comparing behavioral and physiological changes in low and moderately intoxicated users, we aim to gain more accurate  understanding of the effects of cannabis on participants and identify possible behavioral and health risks. Through such in-depth and systematic analysis, we can provide more comprehensive insights on cannabis intoxicated behaviors. In this study, P24 and P63 exhibited different patterns of cannabis use and its effects. P24 reported 12 instances of cannabis use, averaging 4.9 uses per week, with a self-reported feeling of being high averaging only 0.8 (Min=0, Max=6, SD=2.21). This suggests that despite frequent cannabis use, P24 experienced a low level of cannabis intoxication. On the other hand, P63, reported using cannabis only twice during the study period but had a high cannabis intoxication mean of 5.5 (Min=4, Max=7, SD=1.50), indicating a higher level of intoxication per use despite lower frequency. The contrasting usage patterns show the complex variations in individual experiences with cannabis. \textcolor{black}{In our study, by employing a representative sample selection methodology, specifically stratified sample selection based on self-reported levels of cannabis intoxication, we aim to provide insight into the impact of the different dimensions of cannabis use. P24 represents a group of users who experience cannabis at a relatively low intensity (Mean=0.8, SD=1.4) even though they use it frequently (Mean=4.9, SD=5.2  per week), whereas P63 represents a group of users who use cannabis less frequently  (Mean=0.5, SD=0.9  per week) but perceive higher levels of intoxication (Mean=5.5, SD=1.5) each time they use it. This contrast not only reveals the diversity of cannabis use patterns, but also highlights the importance of individual responses to differences in cannabis effects. These two participants were chosen as research examples to demonstrate the potential value of the application of XAI technology in revealing behavioral and physiological changes in cannabis use. Through in-depth analysis of these two common patterns of use, we were able to capture nuances of cannabis use that are critical to understanding how cannabis may affect or is associated with the behavioral and physiological states of individuals under the influence of cannabis use.}

\subsection{Model Building and XAI Analysis}

\begin{figure*}
\centering
\includegraphics[width=1.0\textwidth]{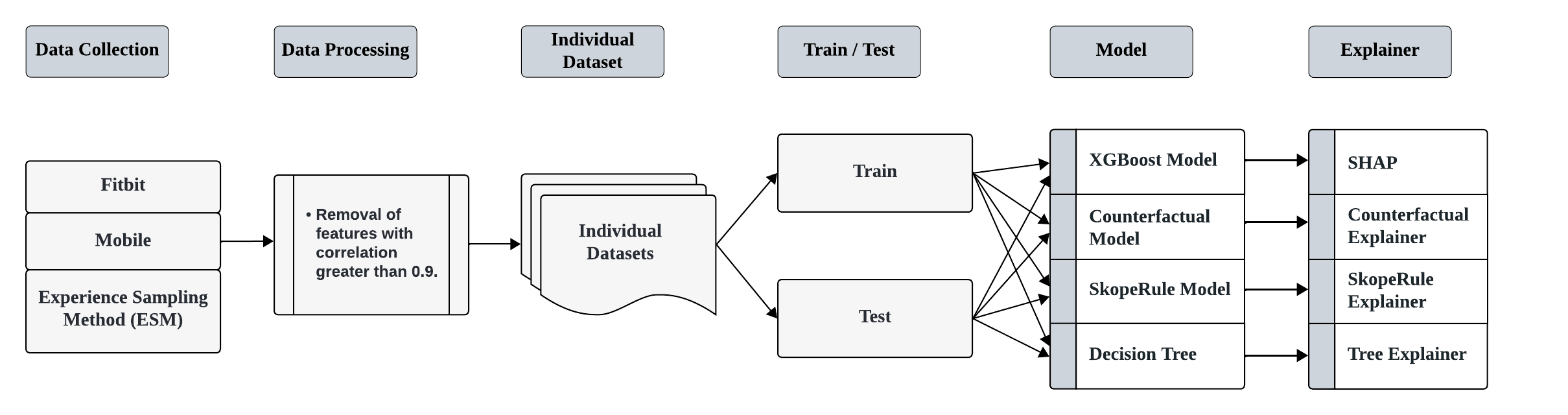}\hfill
\caption{Person-Specific XAI Pipeline}
\label{fig: method-1}
\end{figure*}

% For modeling, we eliminated features with coefficient values greater than 0.9. 
 \textcolor{black}{
 % In performing the feature selection process, we took into account the fact that the features originate from different domains and consequently used a prioritization system based on the type of feature. Features relevant to the use of our Fitbit wearable devices are prioritized above other types of features. Fitbit devices are specifically designed to track health and activity metrics such as steps, heart rate, sleep quality, etc. This data is typically more accurate and comprehensive than data collected by mobile apps because 
% Fitbit devices are constantly worn close to the body and are able to monitor a user's physiology and activity status 24 hours a day. 
We first examined correlations among sensor features in each category. When features exhibited high correlation (coefficient values $>$ 0.9), since heart rate, step count, sleep quality, and a variety of other physiological and activity parameters are more important for cannabis detection, feature belonging to the Fitbit sensor category (e.g., movement) was prioritized and selected. In cases where multiple features of the same feature type were highly correlated, we ensured that the rules were applied consistently (e.g., retaining the higher priority feature), which helped maintain the clarity and repeatability of the process.} We created individual models that were trained on each participant using their own five-minute data segments as the unit of analysis. We imputed missing values with column averages and used MINMAX normalization. For personalized analysis and validation, individual datasets are partitioned from the overall dataset, and data from specific \textcolor{black}{smoking session} is reserved as a test set. Subsequently, multiple machine learning models \textcolor{black}{or XAI packages with machine learning models} (e.g., XGBoost, Decision Trees, SkopeRules, and Counterfactual models) were trained and evaluated to  generate the final analysis results.
%In our modeling approach, we integrate the hyperparametric optimization framework Optuna \cite{optuna_2019} and cross-validation techniques for optimal model tuning. The process starts with Optuna's efficient search for optimal hyperparameters. We use a ten-fold cross-validation approach in Optuna's optimization procedure to verify the performance of different parameter combinations. In this method, the training set is divided into ten subsets. The model is trained on nine subsets and validated on the tenth, and this process is repeated ten times, with each subset used for validation exactly once. After completing the cross-validation cycle, Optuna determines the combination of parameters that produces the best performance, as measured by the F1-score. 
%We then use these optimal parameters to train our model on the entire training set. By using Optuna in conjunction with cross-validation, we can ensure that our models not only perform well on current data, but also generalize efficiently to new, unseen data.}
For testing, we isolated the corresponding data for the \textcolor{black}{smoking session in which} each participant reported cannabis intoxication. We then calculated the average of the data for that \textcolor{black}{smoking session}, and these variables included, but were not limited to, heart rate, number of steps, and location information. By focusing on the average of the \textcolor{black}{smoking session}'s data, we aimed to reduce variability in the physiological and behavioral data, leading to a clearer picture of the effects of cannabis intoxication. The pipleline is shown in Fig. \ref{fig: method-1}

%we focused on the data from individuals who reported cannabis intoxication. We took the one day of the data for the participants who reported cannabis intoxication, calculate the average of this day's data.
%resulting in one line of data per participant on the test dataset Fig. \ref{img:pipline}.

% We categorized the data based on individuals' self-reported use of cannabis that produced euphoria and then calculated the mean of the data for individuals who both smoked cannabis and reported experiencing euphoria. This average was our test set, ensuring that each individual was represented by a single average data point. There is only one line of data for each individual as a test dataset. Its corresponding part is the fourth step of Fig. \ref{img:pipline}.

In order to understand the effects of cannabis use on an individual's behavior and physiology, we analyzed it using a variety of XAI techniques aimed at identifying self-reported patterns of cannabis use and their association with behavioral and physiological changes. 

Since there are many different tools and approaches in the field of XAI, we carefully reviewed and selected a representative analysis tool from each of the major categories of XAI representing feature interaction and importance, knowledge distillation and rule extraction, and intrinsically interpretable models \cite{payrovnaziri2020explainable}. Our selection aimed to ensure the ability to interpret models from different perspectives and dimensions. Interpretation of models, not only adds to the comprehensiveness of the analysis, but also improves our ability to understand the predictions of complex models.

For feature interaction and importance categorization, we used SHAP \cite{lundberg2017unified} to explore how the sensor features influenced the predictions of different individual models, in two different XAI frameworks. More specifically, we used SHAP waterfall plots, where the baseline values denote the model's expected (mean) predictions across the dataset. As the plot progresses, each feature's contribution is sequentially added, ordered by its impact magnitude on the prediction. Red (positive) indicates influence towards the "intoxication" (1) class, whereas blue (negative) indicates influence towards the "not-intoxication" (0) class, conveying the steps from the baseline prediction to the final predicted value.

For the knowledge distillation and rule extraction categorization, we used SkopeRules \cite{skope_rules} to explore the model's internal judgment rules. SkopeRules are first trained on a dataset using a decision tree model. To avoid overfitting and to ensure the diversity of rules, multiple trees are usually trained to form a random forest or a gradient boosting model, while the selected rules are ranked according to their performance metrics.

For the intrinsically interpretable models \cite{loh2011classification} categorization, we used a decision tree model and let the model illustrate how decisions are made. That is, the model generates a graphical representation of a decision tree, showing the structure of the tree, nodes, branches, and decision paths. Based on the nodes, branches and leaves in the model, the function constructs a tree graph. Each node is labeled with a decision rule, and the branches of the tree represent decision paths from parent to child nodes. Blue arrows in the graph represent paths where the condition is satisfied and red arrows represent paths where the condition is not fulfilled. Ultimately, each data point flows through these decision points to the leaf nodes based on its feature values, resulting in the amount of data used to predict intoxication under that decision.

To verify the accuracy of XAI, we used a counterfactual model to modify the top nine most important features in the SHAP results (visible features in the waterfall plots) \cite{yi2023xgboost}. Counterfactual explanations are a method for interpreting predictions made by machine learning models \cite{lim2019these}\cite{mohseni2021multidisciplinary}, that enable users to understand the decision logic of the model and possible directions for improvement by showing possible outcomes under different conditions \cite{shang2022not}. Counterfactual explanations formulate alternate outcomes with minimal adjustments to the input features to identify the key features that change the model's predictions. Counterfactuals explain the behavior of the model, improve user's understanding and trust in the system \cite{wachter2017counterfactual}, and may also help identify potential problems or deficiencies in the model so that it can be improved.

\section{Results}

%After performing SHAP analyses on the machine learning model, we came to the following overall conclusions: Some participants exhibited altered behavior patterns after self-reporting cannabis use. Additionally, a subset of participants use cannabis with effects related to sleep \textcolor{black}{the night before reporting cannabis use.} Lastly, a specific subset of participants was identified who self-reported cannabis use at particular times of the day and days of the week. The distribution of the number of people involved in the SHAP part of the analysis was as follows: the Behavior Changes category consisted of 13 people, the Sleep Effects category consisted of 14 people, and the Time-Specific Use of cannabis category consisted of 7 people. The SHAP analyses provide insight into the behavioral and physiological effects associated with cannabis use in individuals. 
% \textcolor{black}{
% For other analyses, we adopted a representative sample approach. We selected one example participant from each category (e.g., behavior changes), resulting in a total of three participants for analysis. We selected a specific participant from each category as a case study to demonstrate the typical (most prominent and representative) behavioral and physiological characteristics of that category. In this way, we aim to provide a clearer perspective on understanding the effects of cannabis use at the individual-level.
% }

\textcolor{black}{In this section, we present the application of a diverse array of advanced XAI analysis techniques aimed at elucidating the relationship between behavioral and physiological characteristics and self-reported cannabis use for each participant. We elaborate on key findings as follows: 1) Through feature interaction and importance analyses via SHAP, we effectively quantify the influence of interactions among various features on model predictions, 2) Employing knowledge distillation and rule extraction methods via SkopeRules, we derive simple yet insightful rules from intricate models, 3) We explore into intrinsically interpretable models via Tree model, characterized by their transparency, affording us the opportunity to directly scrutinize their internal logic, and 4) Exploring counterfactuals to understand alternative decision options with adjustments of features, thereby enhancing comprehension of the algorithmic decision-making processes associated with cannabis-intoxicated behaviors.}
%Results across these three examples reveal how the techniques from each category work together, providing a more comprehensive view of the acute effects of cannabis use.

% \textcolor{black}{
% In order to effectively communicate our findings and ensure that the key characteristics of each category could be clearly understood, we adopted a representative sample approach. Within each category, we selected a specific participant as a case study as a way of demonstrating the typical behavioral and physiological characteristics of that category. These representative samples were selected on the basis that they exhibited the most prominent and representative characteristics in their respective categories. In this way, we hope to provide a clear perspective on understanding the effects of cannabis use on different individuals.
% }

\subsection{Perturbation-based Feature Importance: SHAP}
% Different Types Of Individuals Who Smoke cannabis Of SHAP}

\subsubsection{Behavior Changes When Smoking Cannabis}

\begin{figure}[htp]
\centering
\includegraphics[width=0.5\textwidth]{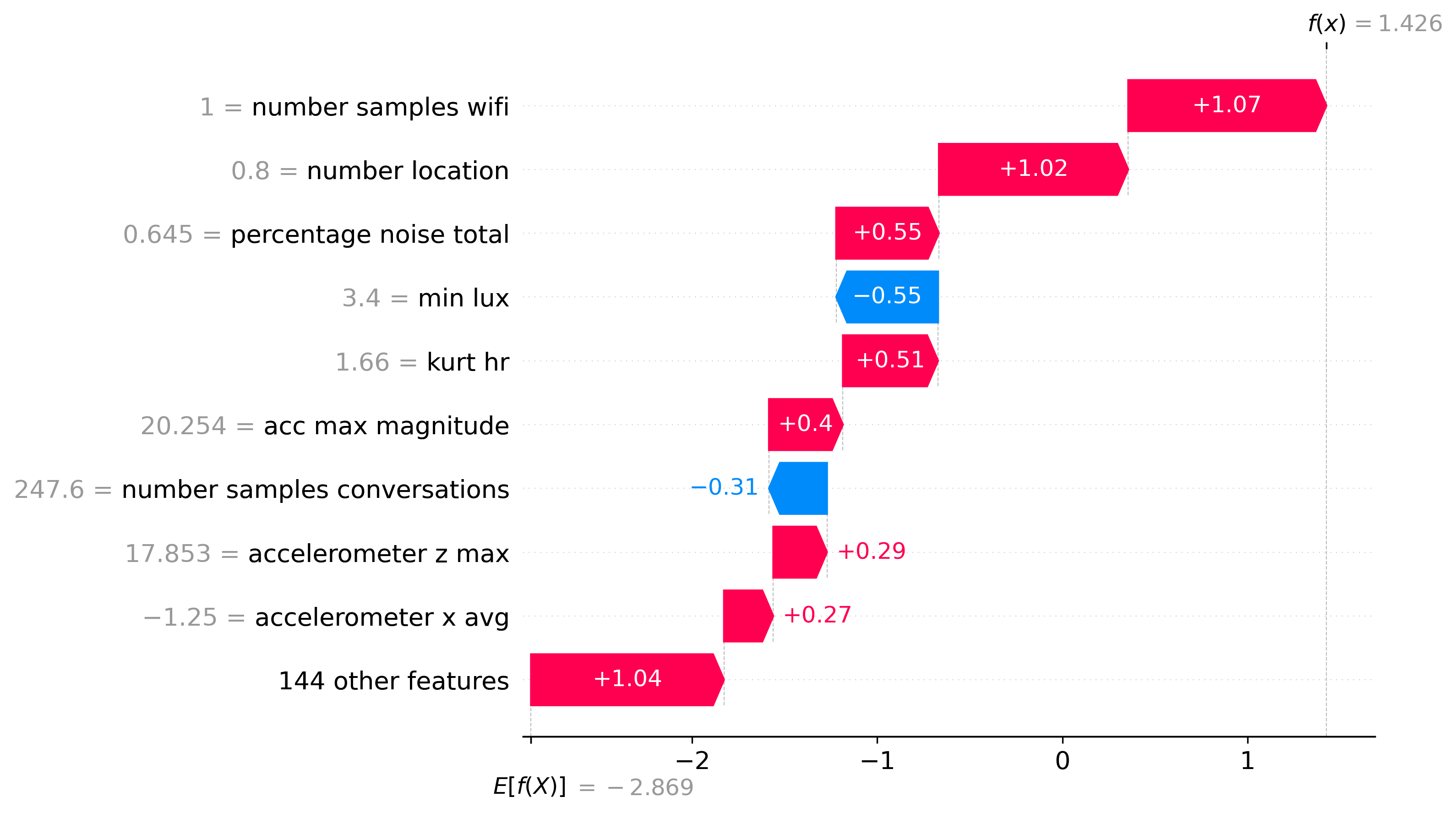}\hfill
% \includegraphics[width=.33\textwidth]{figures/WF_P63_1.png}\hfill
% \includegraphics[width=.33\textwidth]{figures/WF_P65_1.png} \hfill
% \hfill
\caption{SHAP Individual Waterfall Graph (P24)}
\label{fig: shap-1}
\end{figure}
%P63 (Center) and P65 (Right)
\textcolor{black}{SHAP analyses revealed significant behavioral changes in participants during reports of cannabis use. Some participants exhibited increased activity associated with self-reported cannabis use. In these analyses, participants showed frequent changes in activity status, which may indicate that they were more restless or agitated while using cannabis. This restlessness may be reflected in their tendency to use cannabis in public places or noisy environments.}
Looking at data from the example participant for this category, according to Fig. \ref{fig: shap-1}, based on observing the number of Wi-Fi samples a user has, we found that the example participant (P24) tended to remain stationary while using cannabis and connecting to Wi-Fi, usually with only one Wi-Fi connection. Another observation involves the ambient noise around this participant. This participant often experienced a great deal of background noise interference when using cannabis, according to total noise percentage, which accounts for about 60\% of the audio during a call. Such a high percentage of noise may indicate the environment a participant chooses to be in when using cannabis, possibly in public places. Based on the participant's heart rate data, the kurtosis of the heart rates showed that the participant's heart rate was spiking rapidly. At the same time, the data from the accelerometer showed that the participants' phones moved a bit while using cannabis and moved more in the z-axis direction.

\subsubsection{Effects Of Sleep}
\begin{figure}[htp]
\centering
\includegraphics[width=0.5\textwidth]{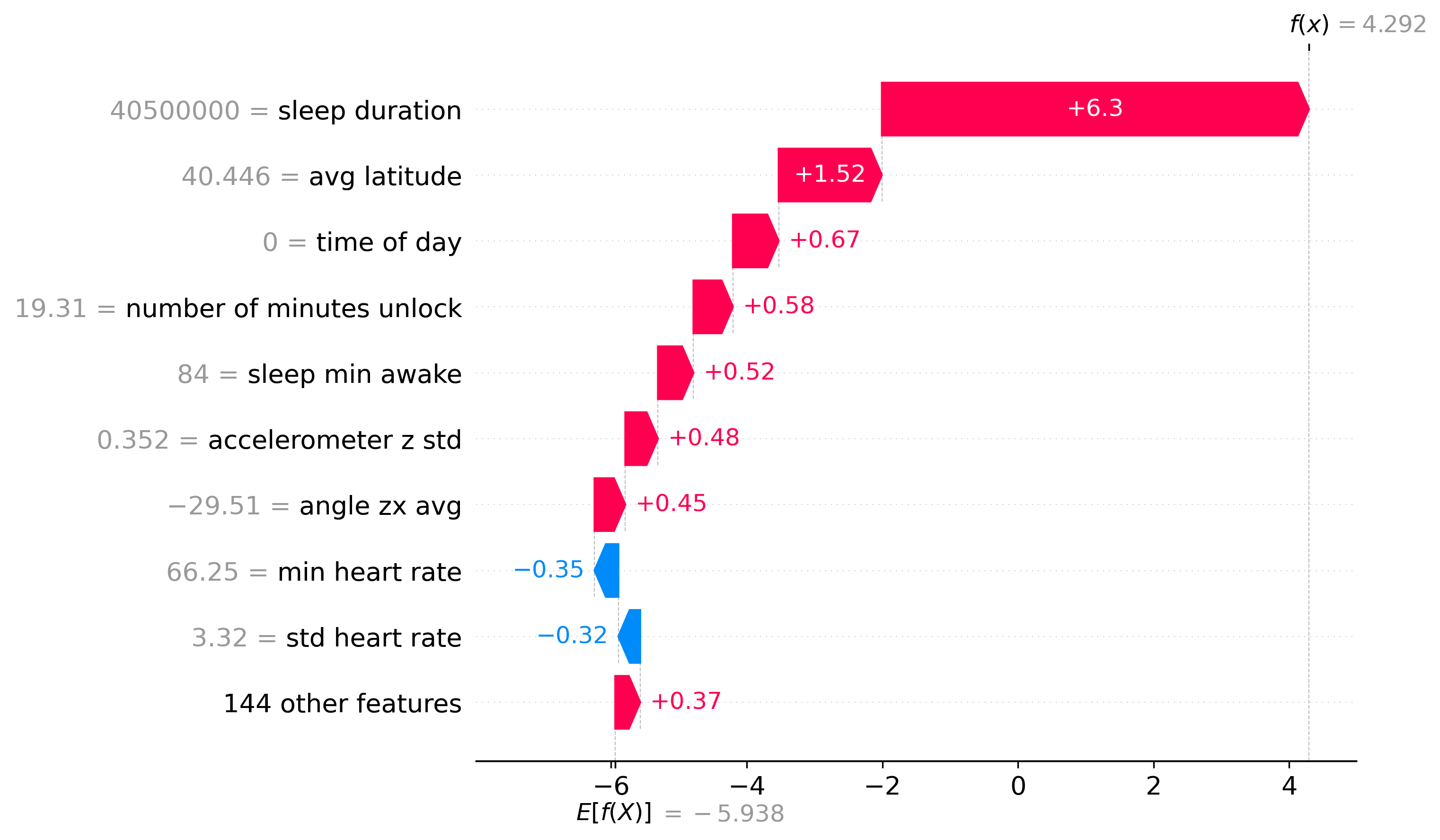}\hfill
% \includegraphics[width=.33\textwidth]{figures/WF_P63_1.png}\hfill
% \includegraphics[width=.33\textwidth]{figures/WF_P65_1.png} \hfill
% \hfill
\caption{SHAP Individual Waterfall Graph (P63)}
\label{fig: shap-2}
\end{figure}
%P63 (Center) and P65 (Right)
% \textcolor{black}{In terms of sleep, SHAP analyses revealed non-traditional sleep patterns in this young adult sample. Many participants had an unusual sleep onset time on the night before cannabis use, usually starting to go to bed after 7:00 pm or before 8:00 pm. This deviation from the regular sleep schedule may be related to recent cannabis use. Although some participants had extended sleep duration, this does not necessarily represent sleep quality or deep rest.}

Examining an example participant’s data, according to Fig. \ref{fig: shap-2}, based on the analysis of sleep data equivalent to over ten hours ($>$ 40 million milliseconds), it was observed that the participant (P63) experienced long sleep duration the night before reporting cannabis use. Time in the early morning hours was positively correlated indicating that the participant may have a habit of using cannabis in the early morning hours. Further, the positive correlation of latitude information with reports of cannabis use indicated that the participant had a relatively regular location for cannabis use. This participant also tended to use their cell phone when using cannabis, based on cell phone lock and unlock times. The participant’s heart data showed a negative correlation with self-reported episode of cannabis use, possibly indicating tolerance to acute cannabis effects on heart rate.

\subsection{Rule-based Explanations: SkopeRules}
% \textcolor{black}{Require a high-level summary}

By applying SkopeRules to sensor data, the generated rule sets reveal diverse patterns of device usage and user behavior. These rules are based on sensor readings such as accelerometer data, light intensity, number of WiFi samples, step speed, heart rate, talk duration and noise level. Results indicated that accelerometer Z-axis data were used to detect device movement or user activity in conjunction with phone brightness to differentiate between activities in different environments; WiFi sample counts were used to analyze network connection status or user location, reflecting user mobility in rich network environments; step speed, acceleration maximal amplitude, and heart rate variability were used to monitor physical activity or health; and location data, call duration, and ambient noise levels were used to analyze participants' social behavior.

In the rule output generated by SkopeRules, each entry contains a rule and three key metrics associated with that rule, which are precision, recall i.e., the rate of correctly recognized by the rule out of all the samples that are actually positive examples. We exemplify the SkopeRules applied in our study by presenting the data of one participant in detail.

\begin{table*}[htbp]
\centering
\caption{Skope Rules (P24)}
\label{tab:scope-rules-1}
\begin{tabular}{|c|c|c|c|}
\hline
PID & Rules & Precision & Recall \\ \hline
P24 & accelerometer x std $>$ 0.04 and min lux $\leq$ 8.0 and number samples wifi $>$ 0.5 & 1.00 & 0.83 \\
\hline
P24 & skew hr $\leq$ 0.79 and kurt hr $>$ -0.36 and number samples wifi $>$ 0.5 & 1.00 & 0.78\\
\hline
P24 & wifi min $\leq$ 2449.5 and accelerometer x avg $\leq$ -1.43 & 0.86 & 0.67\\
\hline
P24 & accelerometer z max $>$ 12.21 and min lux $\leq$ 23.5 & 0.75 & 0.75\\
\hline
P24 & number location $>$ 1.5 and accelerometer y min $\leq$ -0.19 & 1.00 & 0.58\\
\hline
P24 & accelerometer x avg $\leq$ -1.43 and skew dist $\leq$ 0.89 & 1.00 & 0.56\\
\hline
P24 & accelerometer x avg $\leq$ -1.43 and std distance $\leq$ 0.003 & 1.00 & 0.56\\
\hline
P24 & number location $>$ 0.5 and accelerometer x avg $\leq$ -1.43 & 1.00 & 0.50\\
\hline
P24 & number location $>$ 1.5 and acc max magnitude $>$ 12.41 & 1.00 & 0.44\\
\hline
P24 & avg latitude $>$ 20.23 and accelerometer x avg $>$ -1.43 and std steps $>$ 8.31 & 1.00 & 0.33\\
\hline
P24 & accelerometer x avg $>$ -1.43 and accelerometer y avg $\leq$ -0.05 and number of correspondents phone $\leq$ 0.5 & 1.00 & 0.25\\
\hline
P24 & accelerometer x avg $>$ -1.43 and number of correspondents phone $>$ 0.5 & 1.00 & 0.25\\
\hline
P24 & number location $\leq$ 1.5 and accelerometer y avg $\leq$ -0.05 & 1.00 & 0.22\\
\hline
P24 & number location $\leq$ 1.5 and angle yz avg $>$ 90.23 & 1.00 & 0.22\\
\hline
P24 & accelerometer x avg $>$ -1.43 and angle yz avg $>$ 90.30 & 1.00 & 0.22\\
\hline
P24 & accelerometer x avg $>$ -1.43 and angle yz avg $\leq$ 90.30 and number of correspondents phone $>$ 0.5 & 0.50 & 0.11\\
\hline

\end{tabular}
\end{table*}

As shown in Table \ref{tab:scope-rules-1} (P24), the table provides a series of rules based on sensor data (e.g., accelerometer readings, light levels, number of WiFi samples, etc.) for predicting whether an individual uses cannabis. Each rule gives a precision and recall metric for predicting cannabis use based on a specific data threshold. Most of the rules showed very high precision (100\%) but recall varied between 0.11 and 0.83, reflecting the varying effectiveness of these rules in covering all actual cases of cannabis use. Some rules were better able to identify cannabis use, while some rules may have misidentified some cases. The rules include a variety of sensors and data points, such as different axes of accelerometers, light levels, WiFi signal strength, etc., showing attempts to predict cannabis use using a wide range of lifestyle and environmental factors. By combining different sensor data and statistical metrics, the rules are able to accurately predict cannabis use under specific conditions.

\begin{table*}[htbp]
\centering
\caption{Skope Rules (P63)}
\label{tab:scope-rules-2}
\begin{tabular}{|c|c|c|c|}
\hline
PID & Rules & Precision & Recall \\ \hline
P63 & angle yz max $>$ -86.55 and sleep duration $>$ 38310000.0 & 1.00 & 0.57 \\
\hline
P63 &accelerometer x min $\leq$ -0.21 and sleep duration $>$ 38310000.0 & 1.00 & 0.55 \\
\hline
P63 &accelerometer y min $\leq$ -0.15 and sleep duration $>$ 38310000.0 & 1.00 & 0.55 \\
\hline
P63 &accelerometer z min $\leq$ -1.02 and sleep duration $>$ 38310000.0 & 1.00 & 0.53 \\
\hline
P63 &sleep duration $>$ 38310000.0 & 1.00 & 0.50 \\
\hline
P63 &accelerometer z avg $>$ -0.96 and sleep duration $>$ 38310000.0 & 1.00 & 0.50 \\
\hline
P63 &sleep duration $>$ 38310000.0 and acc max magnitude $>$ 1.08 & 1.00 & 0.45 \\
\hline
P63 &sleep duration $>$ 38310000.0 and acc std magnitude $>$ 0.0091 & 1.00 & 0.45 \\
\hline
P63 &min heart rate $\leq$ 25.5 and sleep duration $\leq$ 38310000.0 and day of week $\leq$ 1.5 & 0.56 & 0.56 \\
\hline
P63 &time of day $\leq$ 7.5 and min heart rate $\leq$ 25.5 and sleep duration $\leq$ 38310000.0 & 0.58 & 0.48 \\
\hline
P63 &sleep duration $>$ 38310000.0 and length of conversations seconds $\leq$ 0.96 & 1.00 & 0.33 \\
\hline
P63 &time of day $\leq$ 8.0 and avg latitude $>$ 40.45 and sleep duration $\leq$ 38310000.0 & 1.00 & 0.25 \\
\hline
P63 &avg latitude $>$ 40.45 and sleep duration $\leq$ 38310000.0 and sleep start time $\leq$ 77.5 & 1.00 & 0.24 \\
\hline
P63 &sleep duration $>$ 38310000.0 and length of conversations seconds $>$ 0.96 & 0.67 & 0.11 \\
\hline
\end{tabular}
\end{table*}
As shown in Table \ref{tab:scope-rules-2} (P63),
Based on the SkopeRules result of P63, most of the rules had very high precision, indicating that the predictions were very accurate when these conditions were met. However, recall varied widely, ranging from 0.11 to 0.57, suggesting differences in the ability of some rules to relate to actual cannabis use. Rules that combined sleep duration with other characteristics indicated that participants would have longer sleep durations as well as specific behavioral patterns when using cannabis. The high accuracy of these rules demonstrates their utility in predicting cannabis use, and these rules exemplify the importance of combining diverse data ranging from behavioral changes as well as physiological changes for understanding and predicting the complex behaviors of cannabis use.

\subsection{Tree-based Intrinsically Interpretable Explanations: Decision Tree}

\begin{figure}[htp]
\centering
\includegraphics[width=0.5\textwidth]{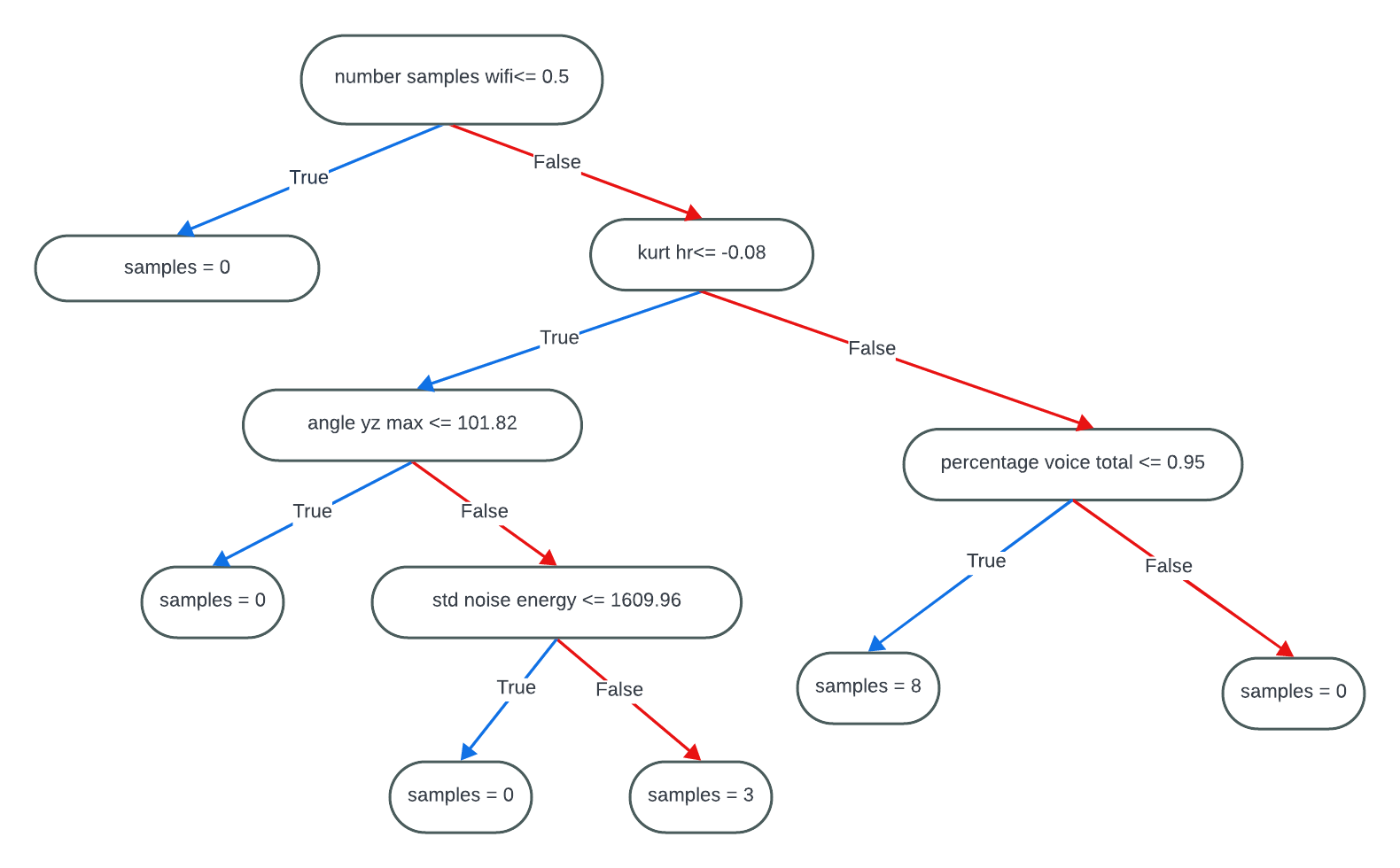}\hfill
\caption{Intrinsically Interpretable Models (P24)}
\label{fig: tree-4}
\end{figure}
We also trained a decision tree classifier following the same methodology as before (see Section III-B).

According to Fig. \ref{fig: tree-4}, based on the root node of the decision tree, we can find that the number of WiFi is a key feature. This indicates that when using cannabis, participants are likely to connect to only one WiFi without moving. If the kurtosis of the heart rate data is less than or equal to -0.08, the model will consider other features. The larger maximum angle of movement of the cell phone in the yz axis combined with the higher standard deviation of ambient noise levels may indicate that the participant may be using cannabis in a public place. When the kurtosis of the heart rate data was greater than -0.08, the model predicted that the participant used cannabis if the percentage of voices in the call was less than or equal to 95\%. This show that there was a certain amount of noise around the participant while using cannabis, which could be due to the fact that they were in a noisy environment.

% \begin{figure}[htp]
% \centering
% \includegraphics[width=0.5\textwidth]{figures/P63-tree-1.png}\hfill
% \caption{Intrinsically interpretable models of P63}
% \label{fig: tree-5}
% \end{figure}

% According to Fig. \ref{fig: tree-5}, this decision tree model predicts whether or not an individual uses cannabis based on characteristics such as sleep patterns, heart rate, duration of daily activities, range of movement, and cell phone usage habits. By calculating the association of shorter sleep duration with minimum wakefulness time, which in turn examines mobility range and cell phone unlocking frequency. The association of longer sleep time with accelerometer Z-axis minimum was further examined for range of movement. The Gini impurity of the leaf nodes shows the purity of each categorization decision, where values close to 0 indicate highly deterministic categorization. Through the combination of these features, the model reveals a complex relationship between cannabis use and specific lifestyle habits.

\begin{figure}[htp]
\centering
\includegraphics[width=0.5\textwidth]{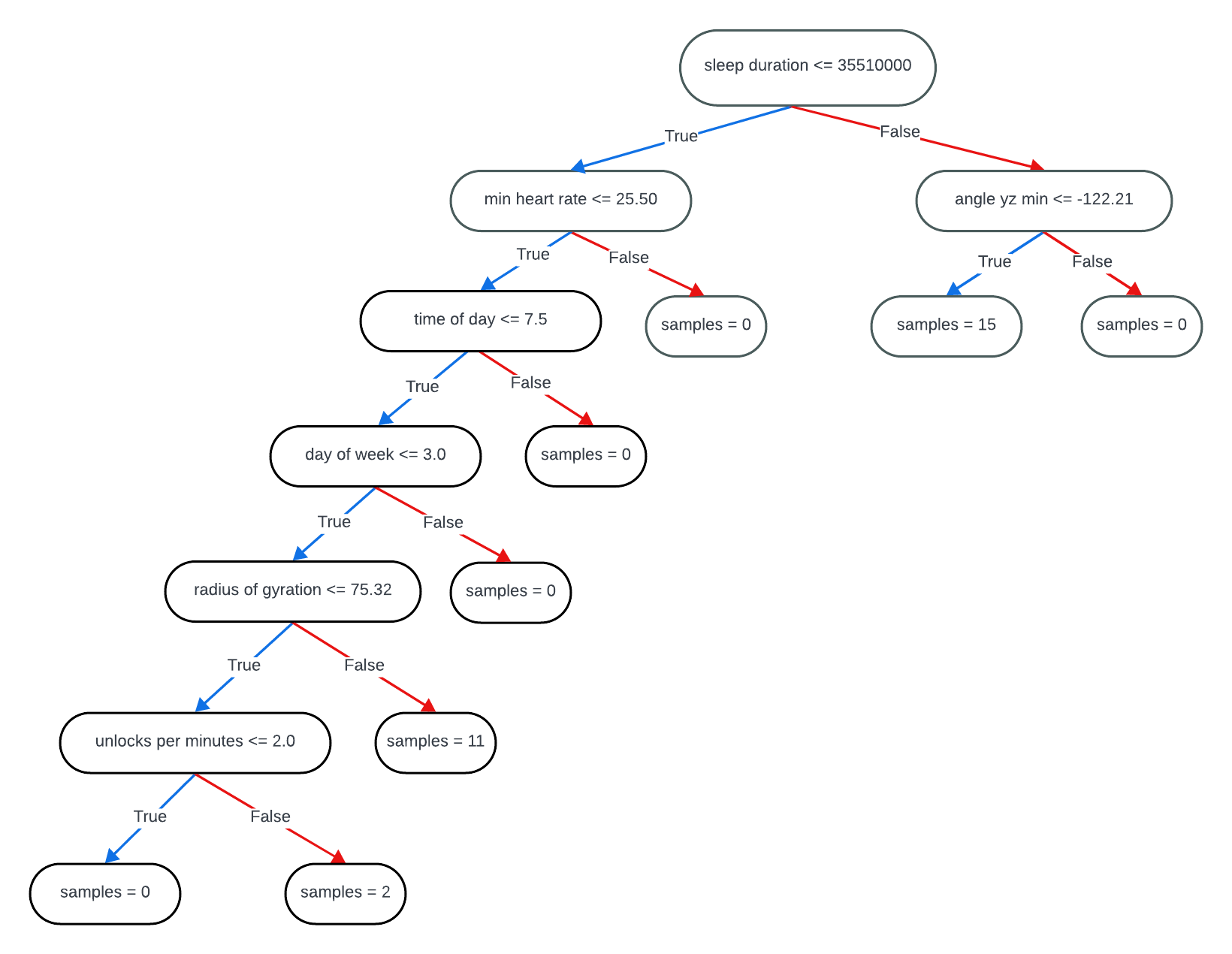}\hfill
\caption{Intrinsically Interpretable Models (P63)}
\label{fig: tree-6}
\end{figure}

According to Fig. \ref{fig: tree-6}, the previous night's sleep duration is the root node, a feature whose units may be milliseconds, and it is used as the primary decision point, indicating its importance in predicting outcomes. Sleep durations less than or equal to approximately 10 hours elicited multiple decision conditions, and participants were highly likely to use cannabis when the angle of movement on the YZ-axis was greater for longer sleep durations.
% Participants use cannabis when sleep duration is less than 10 hours, need, minimum heart rate, time, date, radius of gyration, and number of cell phone unlocks per minute all meet certain conditions.

\subsection{Counterfactual Explanations}

% We conducted a combined analysis using counterfactual explanations to understand the logic of our model's judgments in predicting cannabis intoxication. Based on the counterfactual analysis of 34 local models from 34 participants, we describe the overall patterns observed. Of the nine most important characteristics identified by SHAP, we found that certain indicators were particularly sensitive to self-reported cannabis intoxication. For several participants, deviations in movement patterns altered model predictions, suggesting a strong correlation between movement metrics and self-reported cannabis intoxication. Training of individual models using five-minute snapshots of each participant's data indicated that the perception of intoxication metrics varied considerably across individual models. Some participants required greater adjustments to identified metrics to change the results, suggesting greater resistance or robustness of the model to minor changes. Applying the counterfactual model to this test set again emphasizes the importance of the key characteristics identified by SHAP. Specifically, changes in these characteristics in the average test set resulted in predictable changes in model results, further validating the importance of these metrics in predicting cannabis intoxication. 

\begin{table*}[htbp]
\centering
\caption{Counterfactual explanations (P24)}
\label{tab:cf-1}
\begin{tabular}{|c|c|c|c|c|c|c|c|c|c|c|}
 \hline

 PID &\thead{number samples\\wifi}	&\thead{number\\location}	&\thead{percentage \\noise total}	&min lux	&kurt hr	&\thead{acc max \\magnitude}	&\thead{number samples \\conversations}	&\thead{accelerometer \\z max}	&\thead{accelerometer \\x avg} & label
\\ \hline
P24 & 1.00 & 0.80 & 0.64 & 3.40 & 1.66 & 20.25 & 247.60 & 17.85 & -1.25&1 \\ \hline
P24 & 1.00 & 0.00 & 0.64 & 3.40 & 1.66 & 3.35 & -4.05 & 17.85 & -0.44 &0\\ \hline
P24 & 0.03 & 0.80 & 0.64 & 3.40 & 1.66 & 20.25 & 247.60 & 17.85 & -0.04 &0\\ \hline
P24 & 0.02 & 0.02 & 0.00 & 3.40 & 1.66 & 20.25 & 247.60 & 17.85 & -1.25 &0\\ \hline

\end{tabular}
\end{table*}

% \textcolor{black}{What knowledge did Knowledge Distillation and Rule Extraction provide through the counterfactuals? Based on their benefits, what specific examples of 'Causal reasoning' will be presented, highlighting the 'alternative scenarios' generated by the model when input features are adjusted?}

% Knowledge refinement and rule extraction through counterfactuals can provide insight into the decision-making process of machine learning models by identifying the specific conditions and features that lead to particular predictions. This approach allows for a more nuanced understanding of the model's behavior, especially in areas where the model is complex, such as behavioral predictions related to cannabis use. By generating counterfactual explanations, we can gain insights on how changes in certain input features may have led to different model predictions. In predicting cannabis use, the counterfactual explanations show that certain behaviors or environmental conditions-such as changes in heart rate variability, WiFi sample size, or sleep duration-significantly affect the model's predictions. These examples of counterfactual analysis emphasize the model's ability to generate alternatives based on adjusted input characteristics. This approach not only helps to understand the underlying factors that contribute to model predictions, but also provides valuable insights for intervention by identifying important features that influence outcomes.
Knowledge refinement and rule extraction through counterfactuals can provide insight into the decision-making process of machine learning models by identifying the specific conditions and features that lead to particular predictions.  In predicting cannabis use, the counterfactual explanations show that certain behaviors or environmental conditions-such as changes in heart rate variability, WiFi sample size, or sleep duration-significantly affect the model's predictions. These examples of counterfactual analysis emphasize the model's ability to generate alternatives based on adjusted input characteristics. This approach not only helps to understand the underlying factors that contribute to model predictions, but also provides valuable insights by identifying important features that influence model predicted outcomes.

Based on the counterfactual analysis of Table \ref{tab:cf-1} (P24), the first result demonstrates the potential effects of changes in location and changes in magnitude on cannabis use. The labels changed from 1 to 0 when the number of locations decreased and the acceleration magnitude decreased, suggesting that participants tended not to use cannabis when there was little movement and less activity. The second result showed the effect of "number of wifi samples and acceleration on movement in x. Decreases in the number of wifi samples, as well as decreases in acceleration in the x-axis, were associated with not using cannabis, suggesting that in environments where wifi was not used and there was less physical activity, participants were likely not to use cannabis. The third result shows the effect of cannabis use on location change, noise share, and, number of wifi. Participants tended not to use cannabis when there was less location change and a decrease in noise percentage, and wifi amount. This suggests that participants may be in a public environment when using cannabis.
%we observed that the first and third results were primarily influenced by heart rate data, suggesting that the participant's heart rates kurtosis (kurt) decreases, skewness (skew) increases when using cannabis. The first and third results showed a decrease in the number of WiFi connections, indicating that the participant preferred to stay in a specific place and connected to WiFi when using cannabis. By comparison, the second and third results were more related to the noise feature, that is, when the noise decreased, the results tended to show that the participant did not use cannabis. 

\begin{table*}[htbp]
\centering
\caption{Counterfactual explanations (P63)}
\label{tab:cf-2}
\begin{tabular}{|c|c|c|c|c|c|c|c|c|c|c|}
\hline
PID & sleep duration & avg latitude & \thead{time of\\ day} & \thead{number of\\ minutes unlock} & \thead{sleep min \\awake} & \thead{accelerometer \\z std} & angle zx avg & min heart rate & std heart rate&label \\ \hline
P63 & 40500000.00 & 40.45 & 0.00 & 19.31 & 84.00 & 0.35 & -29.51 & 66.25 & 3.32&1 \\\hline
P63 & -627140.75 & 40.45 & 0.00 & 19.31 & 84.00 & 0.35 & -29.51 & 66.25 & 3.32&0 \\\hline
P63 & -815051.41 & 0.63 & 0.00 & 19.31 & 84.00 & 0.35 & 14.41 & 5.09 & 0.31&0 \\\hline
P63 & -1357805.72 & 40.45 & 0.00 & 19.31 & -12.59 & 0.35 & -29.51 & 66.25 & 3.32&0 \\
\hline

\end{tabular}
\end{table*}
 As shown in Table \ref{tab:cf-2} (P63), the first result showed a decrease in sleep duration, with other indicators remaining almost constant. Participants' sleep the night before had a greater effect on whether or not they used cannabis. The second result shows changes in sleep duration latitude. These changes indicated that participants would use cannabis in specific locations. 
 %all three results showed a change in the angle of the phone or acceleration in the xz-axis, indicating that the pattern of phone use changed when using cannabis. In addition, prior night’s sleep duration was lower, indicating that the person’s use of cannabis might be related to a more fixed location when using cannabis. The third result primarily displayed a decrease in phone unlock time, showing that the participant (P63) spent more time on the cell phone before using cannabis. 

 % Using a variety of XAI tools such as SHAP analysis, rule-based interpretation, intrinsic tree-based interpretable models, and counterfactual interpretation, this analysis provides insights into cannabis use and its behavioral and physiological impacts. The SHAP analysis reveals the importance of specific traits to model predictions, highlighting the impact of environmental conditions and user behavior on cannabis use. Rule-based explanations reveal patterns in device use and user behavior. Decision tree models demonstrate feature-value-based decision paths in an intuitive way, while counterfactual explanations provide ways to understand changes in model predictions by modifying input features.  
 
 In summary, through the use of XAI methods, we have gained a deeper understanding of how cannabis use affects behavior and physiology. This includes changes in behavioral patterns, effects on sleep, and participants' time-specific use patterns. Selecting representative case studies for each category helped to demonstrate the impact of cannabis use on an individual level. By applying feature interaction and importance analyses, knowledge distillation and rule extraction methods, and essentially interpretable models, the study identifies the influence of factors that contribute to the self-reported cannabis use. The XAI methodology used in this study provides unique insights. SHAP analysis focuses on the importance of specific features in predicting behavioral changes resulting from cannabis use, while the knowledge distillation and rule extraction methodology provides clear decision rules describing the conditions under which these changes are likely to occur. Intrinsically interpretable models such as decision trees provide a transparent view of the decision-making process. By analyzing the results of the four XAIs, we can see that several key features emerged across the analyses, factors that are particularly important for understanding cannabis use behavior. In terms of behavioral changes, it was revealed that individuals tend to choose a fixed and stable environment when using cannabis. In terms of physiological changes, changes in heart rate, changes in activity, and changes in sleep duration reflect the effects of cannabis on an individual's physiology and behavior. The emergence and overlap of these key characteristics suggest that cannabis use behavior is the result of a combination of internal and external factors. Individual cannabis use has implications for both physiology and behavior. This complex interaction illustrates the need for a multidimensional, interdisciplinary analytical approach to understanding cannabis use behavior.

\section{Discussion}
Algorithm driven decision making provides a data and model based approach to assessing and predicting individuals' health status. This approach improves the efficiency of clinical decision-making, while combining it with personal models that can help clinicians get insights to develop a treatment plan.

To reveal black-box algorithmic decisions, we utilized a variety of XAI techniques; feature importance (SHAP \cite{lundberg2020local}\cite{lundberg2017unified}), SkopeRules \cite{skope_rules}, intrinsic tree-based explanatory models \cite{scikit-learn}, and counterfactuals \cite{guidotti2022counterfactual} to explore cannabis-intoxicated behaviors.
% the effects of cannabis use on an individual's behavioral and physiological state. 
Each technique provides unique insights that add transparency to the model, enrich physician and patient understanding, and increase the interpretability of the data through intuitive charts or graphs. SHAP analysis provides a granular view of the impact of each trait on the predicted outcome, highlighting key behavioral changes, sleep patterns, and the duration of cannabis use. Through this approach, we can identify specific features that contribute to the observed effects, thereby personalizing the understanding of how cannabis affects the individual. SkopeRules helps to extract interpretable rules that directly link specific sensor data patterns, such as accelerometer readings and sleep duration, to cannabis intoxicated behaviors; physical activity, environmental conditions. This capability offers the potential to obtain actionable insights into how these factors influence a person's behavior and physiological state. Interpretable tree models provide an algorithm's structured decision-making process visualized through decision trees. These models extract complex relationships into easy-to-understand pathways that illustrate how different conditions such as sleep duration and day of the week affect the likelihood of cannabis intoxication. This approach provides a transparent and intuitive way to understand decision logic. Counterfactual explanations help provide insight into the model's sensitivity to certain behaviors and conditions by analyzing how changes in specific characteristics can alter predictions. This analysis highlights key characteristics that have a significant impact on predicting cannabis intoxication, such as sleep duration, cell phone use patterns, and activity levels. Combining these four models provides a more comprehensive understanding of the impact of cannabis, taking advantage of the strengths of each XAI technology to provide a rich, multidimensional analysis, an approach that improves our understanding of the effects of cannabis use on human activities and behaviors.

\subsection{Enhancing Personalized Medicine Through Individual Models and XAI Frameworks}

In our study, by applying different XAI analytics, we explored the acute effects of self-reported cannabis use in terms of behavioral changes, sleep patterns, and duration of use, as well as changes in location. \textcolor{black}{These analyses reveal how cannabis use affects and is associated with an individual's behavioral and physiological state, thus providing more interpretable and transparent data for CDSS to help clinicians make informed decisions. For example, if sleep quality is related to cannabis use or affects sleep quality, interventions could focus on improving sleep hygiene. This tailored approach can improve the effectiveness of treatment programs. Through this approach, healthcare providers can more accurately identify and intervene for specific issues resulting from or derived from cannabis use, thereby promoting the individual's health and recovery.}

Our research achieves enhancements in the analysis process. Not only did we build individual models to gain nuanced insights, but we also explored the context and correlates of cannabis use through detailed interpretable analysis using XAI frameworks. \textcolor{black}{The use of individual models for each participant and the interpretation of the models in conjunction with different XAI techniques represent a major advance in personalizing interventions for substance misuse. This approach reflects the trend in medical research towards personalized medicine and individualized treatment strategies, and emphasizes the importance of tailoring interventions to individual needs and characteristics. While traditional methods provide broad overviews and general analyses, our approach focuses on individual-level data to provide tailored analyses. 
}

\subsection{Practical Applications and Benefits of XAI in Clinical Decision-Making}

\textcolor{black}{
The integration of XAI in CDSS marks a pivotal advancement in healthcare \cite{kunapuli2018decision}\cite{liao2020questioning}\cite{vorm2018assessing}\cite{xie2020chexplain}\cite{lamy2019explainable}\cite{arrieta2020explainable}\cite{schoonderwoerd2021human}\cite{luz2020machine}, particularly in the field of substance misuse treatment \cite{olsson2005one}\cite{schork2019artificial}. According to relevant research, XAI technology enhances trust between physicians and patients by increasing the interpretability of the AI system, enabling physicians to gain a deeper understanding of the model's decision-making process. This interpretability is crucial for CDSS because it helps clinicians better understand and trust the recommendations provided by the system, which in turn increases the likelihood of following them. Particularly in the realm of substance misuse treatment, the in-depth analytics provided by XAI can help physicians more accurately understand the acute impact of cannabis use on an individual's behavioral and physiological state. For example, through different XAI technologies, researchers are able to explore characteristics such as behavioral changes, sleep patterns, and location of use among individuals who report cannabis use. These analyses reveal how cannabis use can alters a person's behavioral patterns and physiological responses, thus providing physicians with information to help develop personalized and targeted interventions. In addition, the application of XAI is not limited to providing transparency of the data, but also includes providing counterfactual interpretations, such as demonstrating possible model predictions under different conditions, which can help physicians and patients better understand the acute effects of cannabis use in a more comprehensive, personalized way, and explore possible directions for improving well-being. Thus, the use of XAI in substance misuse treatment provides clinicians with tools to support more accurate and personalized treatment decisions. With the interpretable data analytics provided by XAI, physicians are better able to understand and respond to the specific needs of their patients, leading to improved treatment outcomes and patient satisfaction.
}
These findings point to the fact that cannabis use may contribute to an individual's propensity to use cannabis in public places or noisy environments, and may also affect sleep quality and duration.

\subsection{Implications for Public Health Research and Future Directions}
\textcolor{black}{We are committed to ethical standards when it comes to research methodology. Our data collection process adhered to strict privacy and ethical guidelines to ensure the confidentiality and security of participants' data. We obtained informed consent from all participants, clearly explaining the purpose of the study and how their data would be used. In addition, we used data anonymization techniques to protect the identity of participants \cite{liu2021digital}\cite{tofighi2018role}\cite{hayes2022using}.}
While focusing on the physiological and behavioral aspects of cannabis use, our study recognizes that the cultural and legal environments surrounding cannabis vary from country to country, and between states in the US. This diversity creates challenges for the applicability of our findings. In regions where cannabis use is legally and culturally accepted, our findings can directly inform public health policy and clinical practice. Conversely, in areas with strict cannabis regulations, our findings may be used more to understand illicit cannabis use and inform intervention strategies. Our study provides a valuable initial contribution in public health research and demonstrates areas for further development \cite{liu2021digital}\cite{tofighi2018role}. \textcolor{black}{Understanding the physiological effects of cannabis use, such as the impact on sleep quality, can guide public health activities, educate about the risks of cannabis use and promote healthier lifestyle choices. Policies and interventions based on these data-driven insights can more effectively address cannabis use in society.}  %The innovative methodology used in analyzing the complex physiological and behavioral responses to cannabis use opens up new avenues for public health inquiry. 
\textcolor{black}{
Ethical considerations ensure that the technology is used for the benefit of patients and does not cause harm. Data security measures protect sensitive patient information and maintain trust in the healthcare system. Accountability mechanisms, such as transparency of the machine learning model decision-making process and the ability to audit and explain machine learning model decisions, reinforce the reliability and integrity of these technologies. Comprehensively addressing these factors will require upholding high standards of ethical principles and protecting patient rights.
}

\subsection{Limitations and Future Work}
\textcolor{black}{
The limited scale of our study necessitates the extension of our methods to larger and more diverse populations. Patterns of cannabis consumption, motivations and physiological responses to cannabis can vary considerably across age groups due to a variety of social, environmental and physiological factors. Therefore, it is important to conduct research across a wide range of age groups to uncover the nuances of how these different populations interact with cannabis. This approach allows for a more detailed understanding of patterns of use, which can inform more targeted public health interventions and policy decisions. Such an extension will provide a more comprehensive understanding of the broad applicability of these techniques in public health research.
}
In addition, our findings highlight the influence of environmental factors on cannabis use. This insight suggests that there is a critical need for future public health strategies and interventions to consider the complex interplay between individual behavior and environmental context. By doing so, we can develop more nuanced and effective approaches to address cannabis related public health issues. We encourage researchers to build on our work and expand research to include different populations and settings. This broader exploration will enrich our knowledge of the complex physiological and behavioral responses to cannabis use and contribute to the development of more effective interventions. 

\textcolor{black}{
We must also acknowledge the potential limitations of XAIs. Reliance on these tools may perpetuate biases inherent in the data they are trained on, leading to biased insights [69]. Healthcare professionals might overly rely on model outputs, potentially diminishing their critical thinking and clinical judgment. Given  the significance of nuances and socio-environmental context in individual cases, it is also important to recognize that our research serves as a tool to complement or augment, rather than  replace, the expertise of clinical decision-making experts .
}

% \subsection{Future Work}
Given the nature of this study, the next steps will focus on gathering empirical evidence to assess the practical utility of XAI visualization for clinicians. A series of structured interviews will be conducted with healthcare professionals to determine their perceptions of the clarity, utility, and educational value of the XAI tool. These interviews will attempt to determine whether the visualizations help in understanding algorithmic decision making and whether XAI enhances trust in AI-assisted decision making. 
\textcolor{black}{
A dedicated working group of interdisciplinary experts, including data scientists, healthcare professionals, ethicists, and patients, could be formed to oversee the development of XAI tools. This will be achieved by aligning the development of XAI tools with the actual needs and challenges faced by healthcare professionals and patients. Incorporating the latest advances in XAI should not just be technical updates, but should also include efforts to address ethical issues, reduce bias, and increase transparency. This will ensure that improvements are not only technologically advanced, but also ethically sound and socially responsible.
}
Additionally, this study will conduct usability testing in which clinicians interact with XAI visualizations in a controlled environment. This next step will gather qualitative data about the user experience, including any challenges or barriers to understanding algorithmic decision making and visualizations. Observations from these sessions will help refine the design of the XAI system to better align with clinicians' cognitive workflows. The overarching goal of future work is to bridge the gap between AI developers in healthcare and end-users to ensure that XAI systems are not only technically sound, but also of real value to the people who rely on them to make critical clinical decisions regarding personalized care.

\section{Conclusion}

In this study, we explored passively "sensed" contexts and correlates of young adult cannabis use. The combined application of SHAP, counterfactual interpretation, rule-based, and intrinsically interpretable model explanations provided an opportunity to gain insight into, and explain, the decision-making logic of the model in predicting self-reported cannabis intoxication. The study found that participants who reported cannabis use exhibit different patterns of use, including signs of regular use, as well as variation in heart rate and sleep patterns that are associated with physiological responses to acute cannabis use. Through the combined application of SHAP and counterfactual interpretation, we were able to identify specific characteristics that had a significant impact on model predictions, such as location data and heart rate, findings that provide important guidance for the development of targeted and personalized health strategies and interventions. Counterfactual modeling analyses complemented the SHAP results, providing a deeper understanding of the model's decision logic. In addition, rule-based and intrinsically interpretable approaches to model interpretation provide decision rules that enable non-specialists to understand how models predict cannabis use based on specific environmental and physiological data. This interpretive enhancement not only improves the transparency of the model, but also strengthens the trust and understanding of the model by healthcare professionals and patients, providing them with an important foundation for developing personalized intervention, when and where it might be needed. 

Our study highlights the critical role of understanding models in interpreting complex health data and providing treatment recommendations. In analyzing the potential effects of cannabis use on sleep patterns and behavioral changes, these models can provide clinicians with the insight needed to develop personalized interventions, which more effectively meet an individual’s needs. Additionally, our approach allows clinicians to optimize diagnostic and treatment decisions using the data analytics provided by XAI technology. By applying XAI technology to clinical practice, we provide physicians with a tool to help them make more informed decisions in line with personalized care.

\section{Acknowledgment}
\textcolor{black}{This study was supported by the National Institute On Drug Abuse of the National Institutes of Health
under Award Numbers U01DA056472 and R21 DA043181. The content is solely the responsibility of the authors and does not
necessarily represent the official views of the National Institutes of Health.}

\bibliographystyle{IEEEtran}
\bibliography{ref}

\end{document}